# A Multi-Path Certification Protocol for Mobile Ad Hoc Networks

Jaydip Sen
Innovation Lab, Tata Consultancy Services Ltd.
Bengal Intelligent Park, Salt Lake Electronic Complex, Kolkata-700091, INDIA
email: Jaydip.Sen@tcs.com

*Abstract*— **A mobile ad hoc network (MANET) is a collection of autonomous nodes that communicate with each other by forming a multi-hop radio network and maintaining connections in a decentralized manner. Security remains a major challenge for these networks due to their features of open medium, dynamically changing topologies, reliance on cooperative algorithms, absence of centralized monitoring points, and lack of clear lines of defense. Most of the routing protocols for MANETs are thus vulnerable to various types of attacks. For security, these protocols are highly dependent on cryptographic key exchange operations. This paper presents a multi-path certification protocol for efficient and reliable key exchange among the nodes in a MANET. Simulation results have shown the effectiveness and efficiency of the protocol.**

## I. INTRODUCTION

A MANET is a collection of wireless hosts that can be rapidly deployed as a multi-hop packet radio network without the aid of any established infrastructure or centralized administrator. Such networks can be used to enable next generation battlefield applications, including situation awareness systems for maneuvering war fighters, and remotely deployed unmanned micro-sensor networks. MANETs have some special characteristic features such as unreliable wireless media (links) used for communication between hosts, constantly changing network topologies and memberships, limited bandwidth, battery, lifetime, and computation power of nodes etc. While these characteristics are essential for the flexibility of MANETs, they introduce specific security concerns that are absent or less severe in wired networks. MANETs are vulnerable to various types of attacks. These include passive eavesdropping, active interfering, impersonation, and denial-of-service. Intrusion prevention measures such as strong authentication and redundant transmission can be used to improve the security of an ad hoc network. However, these techniques can address only a subset of the threats. Moreover, they are costly to implement. The dynamic nature of ad hoc networks requires that prevention techniques should be complemented by detection techniques, which monitor security status of the network and identify malicious behavior.

One of the most critical problems in MANETs is the security vulnerabilities of routing protocols. A set of nodes in a MANET may be compromised in such a way that it may not be possible to detect their malicious behavior easily. Such nodes can generate new routing messages to advertise non-existent links, provide incorrect link state information, and flood other nodes with routing traffic, thus inflicting Byzantine failure in the network. Several secure routing protocols have been proposed for MANETs based on cryptographic mechanisms [1]. Almost all of them assume the existence of a secure channel through which a security association has been established between the source and the destination. However, the prerequisite for such a secure channel to exist is the existence of a security association. This creates a *routing security interdependency cycle* [2].

In this paper, an efficient key exchange protocol is proposed for MANETs that can be easily integrated with a routing protocol thereby providing an integrated framework of routing and security and solving the routing security interdependency cycle. The rest of this paper is organized as follows. Section II presents some related work in MANET security. Section III describes the proposed protocol. Section IV provides performance evaluations of the protocol through simulations. Finally Section V provides concludes the paper.

## II. RELATED WORK

The problem of security and cooperation enhancement among the nodes in a MANET has received considerable attention by the researchers. Cryptography remains the most widely proposed solution to provide authentication nodes. However, cryptography assumes safe key-exchange, which is particularly difficult to realize in multi-hop communications where attackers may launch man-in-the-middle attacks.

Zhou and Haas have introduced a threshold cryptography-based key management scheme for MANETs [3]. A group of n servers together with a master public/private key pair are first deployed by a Certificate Authority (CA). The shares of master private key are generated using threshold cryptography. Thus only $n$ servers together can form the whole signature. If any node wants to join the networks, it must collect all of the *n* partial signatures. This scheme has been extended in a mechanism proposed by Kong et al [4], where a centralized dealer is introduced to issue certificates and private key shares to *t* nodes during the network bootstrapping phase. A threshold cryptography system is deployed in order to provide a (*t*, *n*) secret sharing service. With *statistically unique and cryptographically verifiable* (SUCV) identifiers [5], nodes compute their addresses applying a non-reversible hash function on their public key. Any node can then directly bind a public key (PK) to its owner address and an IP can not be spoofed without the associate private key. Capkun *et al.* [6] proposed a protocol for MANETs by adapting *pretty good privacy* (PGP). In this proposition, a node issues certificates for other nodes. The nodes then exchange certificates and build certification graphs. When two nodes want to exchange their PK, they merge their graphs and establish a certificate chaining in the resulting graph. Eshenaur et al have proposed a trust establishment mechanism in which a node in a MANET can generate trust evidence about any other node. Abdul-Rahman and Hailes have proposed a distributed trust model that uses a recommendation protocol to exchange trust-related information [7]. The model is suited for establishing trust relationships that are less formal and temporary in nature, e.g., some ad hoc commercial transactions.

### III. PROPOSED KEY EXCHANGE PROTOCOL

The proposed protocol integrates a key exchange protocol with routing in a MANET and thus solves the routing-security interdependency cycle [2]. The objective of a routing protocol is to establish a path between a source node and a destination node. To achieve this objective, the reactive routing protocols for MANETs broadcast a *route request* message in the network so that the route to the destination may be discovered. The proposed key exchange protocol utilizes this approach to retrieve the public keys (PKs) of the nodes. To find a certificate of a PK, the source node floods the network with a *certificate request* that is replied either by the target node or by an intermediate node that has a valid certificate of the PK of the target node. The proposed protocol is secure against malicious attackers that may try to distribute spurious certificates in the network and cause routing disruption. To make the protocol robust and reliable, two approaches are taken: (i) multi-path certificate exchange and (ii) trust-based certification. The details of the algorithm are described below.

#### A. Description

In the proposed protocol, it is assumed that every node in a MANET first generates a public/private key pair. Since this key pair is generated by the node itself, the node needs to authenticate with some members in the network before it can join and access network resources. This authentication is based on a certificate exchange. The authentication is mutual. Thus, if a node *S* receives the public key (PK) of a node *D*, *S* issues a certificate for *D*'s PK. In turn, *D* also issues a certificate for *S*'s PK. In the rest of the paper, the set of nodes that has certified for node *S*'s PK is denoted as *K(S)*. As the authentication is mutual, every node in *K(S)* has its PK certified by *S*.

Although the approach of multi-path has not been widely used in certificate exchange schemes for MANETs, it can greatly improve the reliability of a certificate exchange protocol. In designing the proposed protocol two types of multi-path message exchanges are distinguished: (i) multi-path certificate exchange and (ii) multi-path routing. In multi-path certificate exchange, the public key of a node is certified by different nodes (Figure 1 (a)). Due to multiple independent certifications, the confidence assigned to these certificates is higher. A formal computation of the trust values for the certificates may be done using Dempster-Shafer theory [6]. Figure 1 (b) shows an example of a multi-path routing, where a node sends a certificate for another node through multiple node-disjoint paths. Since paths do not have any common node, a malicious node can at most prevent a certificate exchange but cannot spoof the identity of another node during the certificate exchange process.

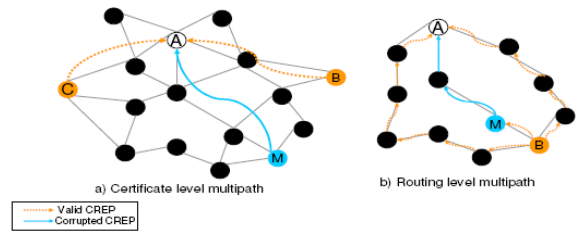

Figure 1. Certificate level multi-path and routing level multi-path

The proposed protocol also utilizes a trust management mechanism [8] to keep track of certification operations. The trust value of a certificate issuing node increases as more number of nodes confirm the public key for which the certificate is issued. On the other hand, when it is detected that a node has issued a spurious certificate, the trust assigned to the node will be decreased and all subsequent certificates issued by the node will also have less confidence associated with them. Consequently, if there is a conflict between certificates, the PK certified by the more trustworthy node(s) will be accepted as genuine.

#### B. Operations

1. *Initialization*: In the proposed protocol, before a node enters the network, it generates a public/private key pair. As a node joining for the first time attempts to get several certificates for of its PK from the existing nodes, it floods the network with a certificate request (CREQ) message.

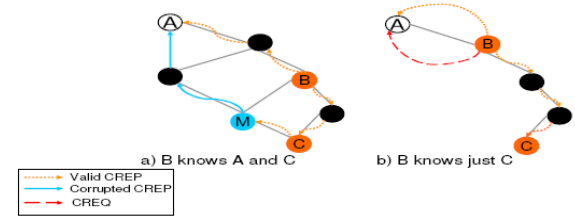

Figure 2. The use of intermediate nodes for certification

2. *Certificate exchange*: Before requesting a node *D*'s certificate, a node *S* evaluates the minimum trust value that is required to consider the public key of *D* as reliable. This threshold value of trust is called the *minimum public key trust value* (MPKTV). This evaluation is local and based on *S*'s security requirements. Node *S* then broadcasts a CREQ for *D*'s certificates including *D*'s address, the list of nodes *K(S)*. The CREQ is sent with a small *time to live* (TTL) to reduce communication overhead of the protocol. Every intermediate node *I* that receives the CREQ checks the PK of *D* and looks in its own certificate list.

If *I* has no certificate for *D* or if it already replied to the CREQ, it simply forwards the packet. Otherwise *I* sends a *certificate reply* (CREP) to *S* containing a certificate of *D*'s public key signed by *I* (Figure 2(a)). If *I* does not know *S*, it constructs a self-signed certificate and informs *S* that it wants to make a certificate exchange (Figure 2(b)). This packet is sent through multiple node-disjoint paths to *S*. If *I* has a route to *D* in its cache, it informs *D* that *S* has requested its PK. *D* responds and requests a certificate for *S*'s PK. Since *I* and *D* can authenticate each other, the communication between *D* and *I* can be made secure by using *I*'s signature. Therefore, no node can corrupt the certificate of *S* issued by *I*. If *D* does not know sufficient number of nodes, it replies to the CREQ itself.

*S* repeats the operation with an increased TTL until it receives the required minimum number of certificates for *D*'s PK. After receiving the certificates, *S* sends the first packet to *D* which includes the list of nodes which has provided the certificates for *D*'s PK. In this way, *D* gets the information about the known certifiers of S. Once they have exchanged their PKs, *S* and *D* issue certificates for each other. This certificate exchange protocol can now be directly applied in routing as S and D do not have to execute any expensive route discovery procedure for routing.

3. *Certificate revocation*: As authentication is mutual, nodes maintain a list of certifiers. An implicit revocation scheme [6] is adopted, where each node periodically updates its public key by communicating secured certificate exchange messages with its peers.

## IV. PERFORMANCE EVALUATION

### A. Simulation parameters

An extensive simulation has been carried out on the proposed scheme to evaluate its performance in several network conditions. The proposed scheme is implemented on network simulator *ns-2*. The simulation was carried out on an abstracted ad network consisting of 100 mobile nodes over an area of 1500 m x 1500 m. The duration of simulation was 120 s. Random way point model has been chosen for node mobility pattern with maximum speed of a node as 10 m/s and average host pause time of 30 s. During each simulation, 5 communications are established that require certificate exchanges among the nodes. The dynamic source routing (DSR) protocol is used for routing [9]. For each configuration, 10 simulations are run and the average value is taken as the result. The following three metrics are studied:

(1) *Valid PK acceptance rate*: it is the ratio of the number of valid public keys accepted and the total number of public keys requested for.

(2) *Corrupted PK acceptance rate*: it is the ratio of the number of corrupted public keys accepted and the total number of public keys requested for.

(3) *Delay*: it is the time interval between the request of a PK and the acceptance of it.

In the simulation, attacks are simulated where the attacker nodes send spurious certificates to the nodes which have requested for those certificates. These attacks can be isolated attacks where every attacker certifies a different PK. However, the attackers may also launch a cooperative attack where different attackers collude and send certifies the same (spurious) PK. Both these types of attacks are simulated. The number of attacker is varied from 0% to 40% of the total number of nodes. Node initialization is not simulated. It is assumed that each node has successfully executed the initialization step by exchanging requisite number of certificates with the honest nodes in the networks. The number of certificate exchange during the initialization is varied from 0 to 20 for each source and destination. A trust value of 0.75 is assigned to a node that is authenticated during the initialization step, while other nodes are assumed to have a trust value of 0.5. MPKTV is varied from 0.5 to 0.9.

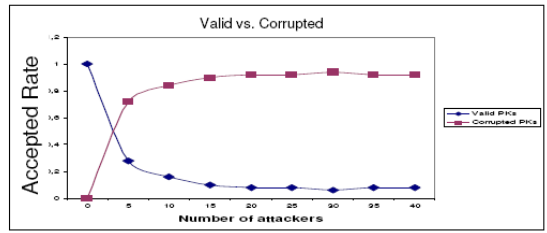

Figure 3. Public key acceptance rate for varying number of attackers

The number of certificate exchanges processed during the initialization goes from 0 to 20 for each source and destination. We assume that a 0.75 trust value is assigned to node authenticated during the initialization step, while 0.5 is assigned to other nodes. The MPKTV varies form 0.5 to 0.9.

### B. Analysis

1. *Isolated attackers*: Figure 3 depicts the variation of the valid PK acceptance rate and the corrupted PK acceptance rate with varying number of attackers. The MPKTV is kept constant at 0.5 while the number of attackers is varied from 0 to 40. There was no initial trust between any pair of nodes in the networks. It is observed that the rate of valid PK acceptance falls rapidly as the number of attackers increases. The trend is just the reverse for the rate of corrupted PK acceptance. Since there was no initial trust at the initial stage, no intermediate node could issue a certificate for a requesting node. Only the destination node could reply to a CREQ message. When there are more attacker nodes in the network, there is a higher probability that an attacker sends a reply to a CREQ message. Since MPKTV is taken 0.5, every PK is considered as valid and accepted by the requester. For any higher value of MPKTV, no PK will be accepted since no node has enough level of trust for issuing an acceptable certificate.

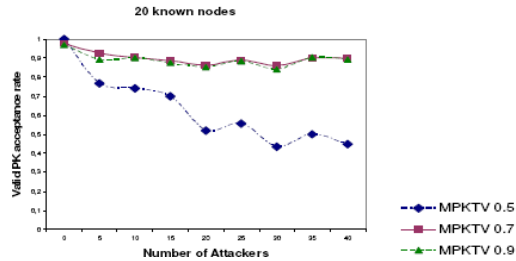

Figure 4. The acceptance rate of valid PK with varying no. of attackers

Figure 4 shows the valid PK acceptance rate with initially 20 known nodes in the network for different values of MPKTV. It is observed that except when MPKTV = 0.5, the acceptance rate is as high as 90%. It has also been observed that the acceptance rate of valid PK increases by 10% when the number of initially known nodes is increased form 5 to 10 and then from 10 to 20 (the figures for the cases of initially known nodes 5 and 10 are not produced to space constraint). With more number of nodes initially known, more nodes send replies to a CREQ message, the average trust in a CREP message increases and thus more PKs are accepted. When MPKTV value is 0.5, any reply is accepted, and the probability to receive a valid PK increases. As the number of attackers increases and becomes more than the number of nodes initially known, the probability of accepting corrupted PKs increases.

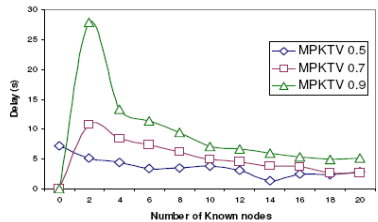

Figure 5. Delay is certification for varying number of known nodes

Figure 5 shows the delay associated in receiving the reply to a CREQ in absence of any attacker node in the network. As the number of known nodes increases, the time required to receive a sufficient number of replies decreases to satisfy a given MPKTV. Moreover, the delay increases with MPKTV because of the increasing requirement for acceptance of a PK. However, a node that has many certifiers of its PK will be quickly authenticated even when the MPKTV is high. This is validated in Figure 5 as delay is found to decrease with increase in number of nodes known initially in the network.

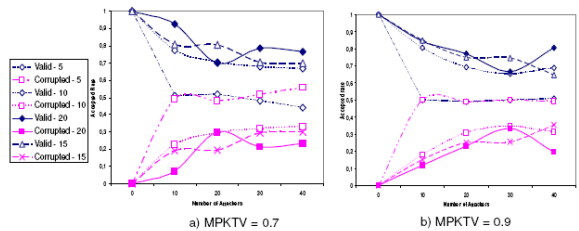

Figure 6. Performance in MANET attacked by a group of colluding nodes (MPKTV= 0.7 in (a) and 0.9 in (b))

2. *Colluding attackers*: Finally, the proposed protocol is simulated in a scenario with colluding attacker nodes. In Figure 6 (a) and Figure 6 (b), valid-$x$ and corrupted-$x$ denote the rate of acceptance of valid and corrupted certificates respectively when $x$ nodes are known initially in the network. It may be observed that some corrupted PKs are accepted since a sufficient number of colluding attackers issue certificates for these corrupted PKs. With MPKTV = 0.9 the rate of acceptance of corrupted PKs is less but as expected, it increases with the number of attackers. Similarly, the rate of acceptance of valid certificates increases with the increase in the known nodes in the network. Nevertheless the increases are much more important when the number of known nodes goes from 5 to 10 that for other increases. An interesting point to note is that with 20 nodes known to the source and the destination, the acceptance rate of corrupted certificates decreases even when the number of attackers increases from 30 to 40. This is because of the fact that when the network contains many attackers, source and destination are more likely to have known nodes in common. Since the common nodes are safe certifiers, with more common nodes in the network, higher is the probability of safe certificate exchange.

## V. CONCLUSION

In this paper, a key exchange protocol for MANETs is proposed that can be integrated with a routing protocol. The protocol is light-weight, efficient and alleviates the routing-security interdependency cycle. Simulation results show that the protocol is resistant to isolated attack launched by malicious nodes that may introduce spurious certificates in the networks. It also performs well against cooperative attacks when sufficient level of trust exists among some nodes before the network deployment.